\documentclass[aip,apl,reprint,groupedaddress]{revtex4-1}

\usepackage{graphicx}
\usepackage{dcolumn}
\usepackage{bm}
\usepackage{amsmath}

\begin{document}

\title{Band offsets of semiconductor heterostructures: a hybrid density functional study}

\author{Amita Wadehra}
\email[Correspondence author; email: ]{amita@mps.ohio-state.edu}
\affiliation{Department of Physics, The Ohio State University, Columbus, Ohio 43210, USA}

\author{Jeremy W.~Nicklas}
\affiliation{Department of Physics, The Ohio State University, Columbus, Ohio 43210, USA}

\author{John W.~Wilkins}
\affiliation{Department of Physics, The Ohio State University, Columbus, Ohio 43210, USA}

\date{\today}

\begin{abstract}
  We demonstrate the accuracy of the hybrid functional HSE06 for computing
  band offsets of semiconductor alloy heterostructures. The highlight
  of this study is the computation of conduction band offsets with a
  reliability that has eluded standard density functional theory. A
  high-quality special quasirandom structure models an infinite
  random pseudobinary alloy for constructing heterostructures along
  the (001) growth direction. Our excellent results for a variety of
  heterostructures establish HSE06's relevance to band engineering of
  high-performance electrical and optoelectronic devices.

\end{abstract}


\maketitle

Heterostructures are ubiquitous in semiconductor technology. For
instance, AlInAs/InGaAs is used for quantum cascade lasers and
infrared photodetectors; InGaP/AlGaAs for high electron mobility
transistors (HEMTs), heterojunction bipolar transistors (HBTs),
and phototransistors; AlInAs/InP for HEMTs; InGaP/GaAs for HBTs; and
InGaAs/InP for single-photon avalanche photodiodes and HBTs. Among
the most important properties that determine the feasibility and
performance of heterostructure devices are the band offsets. These are
the discontinuities between the valence band maxima (VBM) or conduction
bands minima (CBM) of each semiconductor at their common interface,
and act as barriers to electrical transport across the interface. Band
engineering of novel devices with desired properties, particularly
quantum cascade lasers and quantum dot-based devices, critically require
a precise knowledge of band offsets. However, reliable measurements and
predictions of band offsets continue to be challenging despite extensive
theoretical and experimental efforts.~\cite{Yu1992, Vurgaftman2001, Adachi2009}

Density functional theory (DFT) is an efficient method for calculating
electronic structure. The accuracy of DFT calculations is controlled
by the choice of exchange-correlation (XC) functional. Local and
semi-local functionals such as LDA (local density approximation) and PBE
(Perdew-Burke-Ernzerhof)~\cite{Perdew1996} fail to produce accurate
bandgaps, and in extreme cases predict small gap semiconductors
as metals. Hybrid XC functionals, that include a fraction of
Hartree-Fock (HF) exchange, provide a promising alternative. In this
letter, we determine the suitability of a hybrid functional HSE06
(Heyd-Scuseria-Ernzerhof)~\cite{Heyd2003} to compute band offsets of
several III-V compounds and pseudobinary alloy heterostructures. HSE06
includes a fraction, $\alpha$, of screened, short-range HF exchange
to improve the derivative discontinuity of the Kohn-Sham potential
for integer electron numbers (default HSE06 uses $\alpha$=0.25). This
functional was recently used to predict the band alignments throughout
the composition range of InGaN.~\cite{Moses2010} However, the bands
were constructed and aligned with respect to vacuum instead of the
conventional method of computing the band alignment involving a
heterostructure supercell.

\begin{figure}
\includegraphics{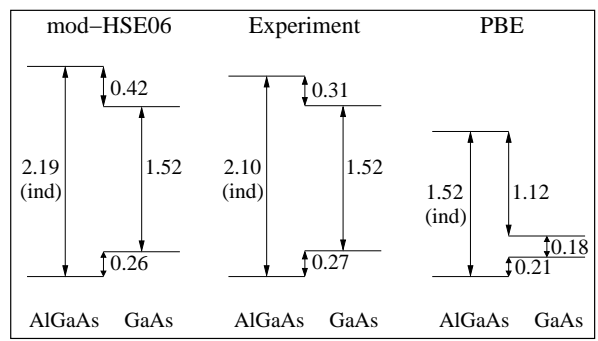}
\caption{\label{fig:algaas}Band alignments for Al$_{0.5}$Ga$_{0.5}$As/GaAs
heterostructure computed with mod-HSE06 ($\alpha=0.30$, see text)
and PBE, in comparison with experiment.~\cite{Vurgaftman2001} The
direct and indirect bandgaps are shown for GaAs and Al$_{0.5}$Ga$_{0.5}$As, respectively. The hybrid functional shows
significant improvement over PBE for bulk bandgaps and both valence
and conduction band offsets of the heterostructure.} \end{figure}

Figure~\ref{fig:algaas} highlights the success of HSE06 in
computing valence and conduction band offsets of the classic
Al$_{0.5}$Ga$_{0.5}$As/GaAs heterostructure in close agreement with
experiment. HSE06 also shows significant improvement over PBE for
computing accurate lattice constants, bandgaps, and cation outermost
\textit{d}-orbital binding energies for the bulk III-V compound
semiconductors.~\cite{Wadehra2010} These results signal advantages
of HSE06 over traditional functionals for computing electronic properties.

Since the percentage of HF exchange in a hybrid functional is not a
universal constant and the optimal value may be system-dependent, it is
worthwhile to study the variation in bandgaps as a function of $\alpha$
in HSE06. Figure 2 demonstrates close agreement between the computed
and experimental direct bandgaps for III-V phosphides and arsenides
using the default HSE06 functional. Since $\alpha = 0.30$ describes the
bandgaps of both AlAs and GaAs so well, a reasonable choice for $\alpha$
in the AlGaAs alloy would be 0.30, which we refer to as mod-HSE06. For
all other systems we use the default $\alpha = 0.25$ as no other value
will work for both components of the alloy.

\begin{figure}
\begin{center}
\includegraphics{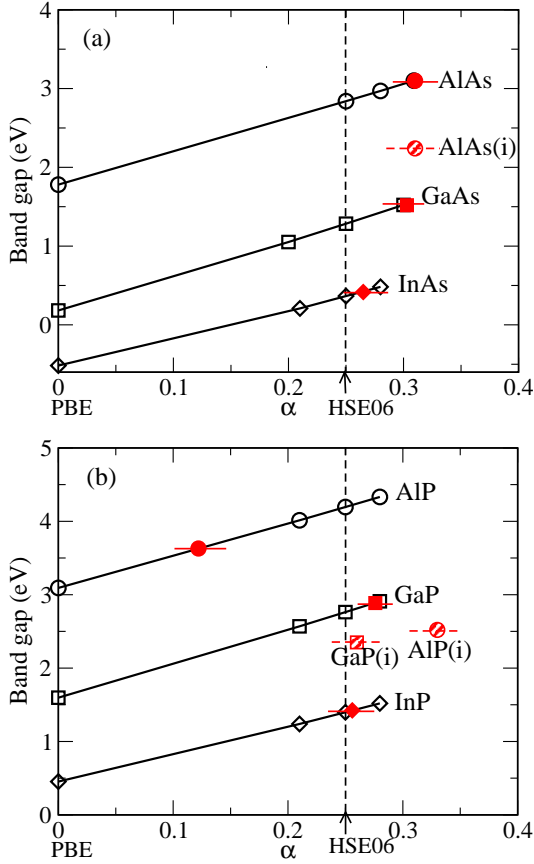}
\caption{\label{fig:bandgaps}(Color online) Calculated direct bandgaps
(open symbols) vs. fraction of Hartree-Fock mixing ($\alpha$) in HSE06
functional for III-V (a) arsenides and (b) phosphides. The vertical line
passes through $\alpha$=0.25, the default fraction for HSE06. PBE results
are shown at $\alpha$=0. Filled (partially filled) red symbols indicate
direct (indirect) experimental bandgaps. The experimental
bandgaps do not vary with $\alpha$ but are positioned according to the
$\alpha$ needed in HSE06 to reproduce those values. Default HSE06 gives
bandgaps close to experimental values for InAs, GaP and InP. An
optimal value of $\alpha$=0.3 is required to obtain the experimental gaps for
both AlAs and GaAs. We use default HSE06 for AlP as different $\alpha$
values of 0.33 and 0.12 are required to match experimental indirect and
direct bandgaps, respectively.}

\end{center}
\end{figure}

We employ the average electrostatic potential
technique~\cite{Baldereschi1988} to compute the band offsets of
the heterostructures S1/S2, where S1 and S2 are the semiconductors
constructing a heterostructure.  The bulk valence band edges are aligned
through a reference potential calculated across the interface of the
heterostructure. The difference in valence band maxima of S2 and S1
is $\Delta E_{\textrm{VBM}}$. The discontinuity in this reference
potential across the heterostructure interface is defined as $\Delta
V_{\textrm{step}}$. The valence band offset, $\Delta E_{\textrm{v}}$,
is calculated as

\begin{equation}
\Delta E_{\textrm{v}}^{\textrm{(S2-S1)}} = \Delta E_{\textrm{VBM}}^{\textrm{(S2-S1)}} + \Delta V_{\textrm{step}}^{\textrm{(S2-S1)}}
\label{eq:vbo}
\end{equation}

\noindent The conduction band offset is determined from $\Delta
E_{\textrm{v}}$ and the difference in bulk bandgaps, $\Delta
E_{\textrm{g}}$, as

\begin{equation}
\Delta E_{\textrm{c}}^{\textrm{(S2-S1)}} = \Delta E_{\textrm{g}}^{\textrm{(S2-S1)}} + \Delta E_{\textrm{v}}^{\textrm{(S2-S1)}}
\label{eq:cbo}
\end{equation}

\begin{table}\footnotesize
\begin{ruledtabular}
\renewcommand{\arraystretch}{1.3}
\newcolumntype{Z}{>{\centering\arraybackslash}X}
\caption{\label{tab:offsets} Comparison of HSE06 and PBE calculated
valence band offsets $\Delta E_{\textrm{v}}$ (eV) and conduction band
offsets $\Delta E_{\textrm{c}}$ (eV) of III-V binary and pseudobinary
alloy heterostructures with experiment values. All the alloys in the
present study are of the form A$_{0.5}$B$_{0.5}$C. The band offsets
computed with HSE06 show much better agreement with experimental values
than PBE. Asterisks indicate that mod-HSE06 ($\alpha$=0.30) is used
instead of the default HSE06 functional ($\alpha$=0.25). A positive value
of $\Delta E_{\textrm{v}}$ ($\Delta E_{\textrm{c}}$) for a heterostructure
S1/S2 implies that the valence (conduction) band edge of semiconductor S2
lies higher than that of the semiconductor S1. The valence and conduction
band offsets have opposite signs for type-I (straddling) heterostructures
and the same sign for type-II (staggered) heterostructures.}

\begin{tabular*}{\textwidth}{l @{\extracolsep{\fill}} ccccccc}
Heterostructure & \multicolumn{3}{c}{$\Delta E_{\textrm{v}}$ (eV)} & & \multicolumn{3}{c}{$\Delta E_{\textrm{c}}$ (eV)}\\\cline{2-4} \cline{6-8}S1/S2    & HSE06 & PBE & Exp. & & HSE06 & PBE & Exp.\\
\hline

AlAs/GaAs & 0.52* & 0.45 & 0.53\footnotemark[1] & & -1.02* & -1.08 & -1.05\footnotemark[1]\\

AlP/GaP\footnotemark[2] & 0.54 & 0.49 & 0.55\footnotemark[3] & & 0.58 & 0.57 & 0.38\footnotemark[3]\\

AlSb/GaSb\footnotemark[2] & 0.38 & 0.35 & 0.38\footnotemark[1] & & -0.67 & -1.21 & -0.51\footnotemark[4] \\
\hline

AlGaAs/GaAs\footnotemark[2] & 0.26* & 0.21 & 0.27\footnotemark[1] & & -0.42* & -1.12 & -0.31\footnotemark[1]\\

InGaP/GaAs & 0.32 & 0.26 & 0.31\footnotemark[1] & & -0.24 & -0.34 & -0.18\footnotemark[1]\\

InP/InGaAs & 0.36 & 0.27 & 0.34\footnotemark[5] & & -0.38 & -0.42 & -0.27\footnotemark[5]\\

InP/AlInAs & 0.16 & 0.14 & 0.17\footnotemark[6] & & 0.22 & 0.17 & 0.25\footnotemark[6]\\
\hline

AlInP/InGaP\footnotemark[2] & 0.22 & 0.19 & 0.24\footnotemark[7] & & -0.23 & -0.74 & -0.26\footnotemark[7]\\

AlInAs/InGaAs & 0.23 & 0.18 & 0.22\footnotemark[8] & & -0.57 & -0.54 & -0.51\footnotemark[8]\\

InGaP/AlGaAs & 0.11 & 0.08 & 0.09\footnotemark[9] & & 0.25 & 0.19 & 0.28\footnotemark[10]\\
\hline

d/o InGaP\footnotemark[11] & 0.01 & 0.02  & & & -0.24 & -0.18 & 0.15\footnotemark[12]\\

\end{tabular*}
\end{ruledtabular}
\footnotetext[1]{From Ref.~\cite{Vurgaftman2001} and references therein}
\footnotetext[2]{The indirect bandgaps of AlP, GaP, AlSb, AlGaAs, AlInP
are used to get $\Delta E_{\textrm{c}}$ for these heterostructures.}
\footnotetext[3]{From Ref.~\cite{Nagao1997}.}
\footnotetext[4]{No exp. data; $\Delta E_{\textrm{c}}$
calculated from exp. $\Delta E_{\textrm{v}}$ and $\Delta E_{\textrm{g}}$}
\footnotetext[5]{From Ref.~\cite{Waldrop1991}.}
\footnotetext[6]{From Ref.~\cite{Bohrer1993}.}
\footnotetext[7]{From Ref.~\cite{Patel1993}.}
\footnotetext[8]{From Ref.~\cite{Adachi2009} and references therein}
\footnotetext[9]{No exp. data; $\Delta E_{\textrm{v}}$
calculated from exp. $\Delta E_{\textrm{c}}$ and $\Delta E_{\textrm{g}}$}
\footnotetext[10]{From Ref.~\cite{Kim1995}.}
\footnotetext[11]{d=disordered and o=ordered with Cu-Pt (L$1_{1}$) ordering}
\footnotetext[12]{From Ref.~\cite{Schneider1994}.}
\end{table}

Table~\ref{tab:offsets}, by comparing the DFT and experimental
band offsets for several III-V compounds and pseudobinary alloy
heterostructures, establishes that both conduction and valence
band offsets calculated by HSE06 are much closer to experiments
than PBE. HSE06 predicts the accurate nature and magnitude of the
bandgaps~\cite{Wadehra2010} and hence band offsets, whereas for PBE, the
error in bandgaps translates to error in conduction band offsets. Sampling
Table~\ref{tab:offsets}, we start with AlSb/GaSb. PBE underestimates the
bandgaps of both constituents,~\cite{Wadehra2010} predicts GaSb metallic,
and produces a giant $\Delta E_{\textrm{c}}$. HSE06 corrects the bandgaps
as well as offsets. A major success of HSE06 is evident for the more
complex alloy heterostructures such as AlInP/InGaP; HSE06 predicts
AlInP as indirect bandgap material, in agreement with experiment, and
computes accurate offsets. PBE, on the other hand, predicts AlInP as a
direct gap semiconductor with a bandgap of 1.52 eV (and indirect gap of
1.72 eV).~\cite{Wadehra2010} The PBE calculated $\Delta E_{\textrm{c}}$,
using the direct (indirect) bandgap, is -0.54 eV (-0.74 eV), much bigger
than the experimental value. Another specific example is the semiconductor
InGaAs, which PBE predicts as metallic.~\cite{Wadehra2010} Although both
PBE and HSE06 give numerically similar band offsets for AlInAs/InGaAs
and InP/InGaAs, PBE fails on the physics. A recent study on pseudobinary
alloys shows that HSE06 accurately predicts the alloy concentration for
the direct-indirect bandgap crossovers.~\cite{Nicklas2010} These results
nail HSE06's predictive power for optoelectronics.

We use the planewave projector augmented-wave (PAW)
method~\cite{Blochl1994} with PBE and the HSE06 hybrid functional in
the \textsc{vasp} code.~\cite{Kresse1996, Kresse1999, Paier2006} The
outermost \textit{d} electrons of cations are treated as valence. We
use a planewave energy cut-off of 500 eV for all our calculations. The
Brillouin zone integration for the bulk III-V binaries and their
heterostructures is performed on a $\Gamma$-centered 8x8x8 and 8x8x1
k-point meshes, respectively. For the pseudobinary alloys and alloy
heterostructures, 4x8x4 and 4x8x2 $\Gamma$-centered k-point meshes are
used, respectively.  Atomic relaxations are not taken into account for the
common-anion materials. However, when the anions are different in the two
semiconductors making a heterostructure, we relax the interfacial atoms.

The computation of an infinite random alloy requires a large supercell
and is arduous, particularly with the hybrid functional. In order
to simulate such an alloy using a finite supercell with reasonable
computational effort, we employ a special quasirandom structure
(SQS),~\cite{Zunger1990} generated using the Alloy Theoretic Automated
Toolkit (ATAT).~\cite{Walle2002} Only the 16 cations on a fcc sublattice
were distributed according to the SQS construction, whereas the 16
anions are located on the separate sublattice that makes up the 32 atom
zincblende supercell.~\cite{Wadehra2010} The degree to which this SQS
matches an infinite perfect random alloy is based on the behavior of the
first few radial correlation functions. We search all possible 16-atom
fcc supercells with two lattice-vectors orthogonal to (001) for easy
construction of a heterostructure in a (001) growth direction yielding the
smallest deviation in the radial correlation functions from the perfectly
random alloy. The SQS employed in this work has radial correlation
functions that match the perfect random alloy up to the fourth nearest
neighbor pairs. The substrate lattice constant, computed with either
functional, is used for alloy heterostructures. The heterostructures
of alloys (binary III-V's) are modeled as 4+4-layer thick supercells,
64 atoms (16 atoms) with a (001) interface. For either functional,
the average of the lattice constants of the nearly lattice-matched
bulk materials, computed with that particular functional, is used for
the heterostructure.

To conclude, we have accurately computed the valence and essentially the
conduction band offsets, using the hybrid functional HSE06, for several
heterostructures of direct technological significance. The importance of
this study lies in its overcoming the limitations of theory for excited
states that restricted its application to semiconductor electronics. Our
excellent results for a broad range of materials indicate that theory
can predict structures with desirable electrical and optoelectronic
properties. More resources may conquer strained and lattice-mismatched
interfaces increasingly being used for advanced devices.

This work was supported by DOE-BES-DMS (DE-FG02-99ER45795). We
used computational resources of the NERSC, supported by the U.S. DOE
(DE-AC02-05CH11231), and the OSC. We thank Steven A. Ringel, Siddharth
Rajan, and Richard G. Hennig for useful suggestions, and Georg Kresse
for the beta version, \textsc{vasp}5.1.


\begin{thebibliography}{22}
\expandafter\ifx\csname natexlab\endcsname\relax\def\natexlab#1{#1}\fi
\expandafter\ifx\csname bibnamefont\endcsname\relax
  \def\bibnamefont#1{#1}\fi
\expandafter\ifx\csname bibfnamefont\endcsname\relax
  \def\bibfnamefont#1{#1}\fi
\expandafter\ifx\csname citenamefont\endcsname\relax
  \def\citenamefont#1{#1}\fi
\expandafter\ifx\csname url\endcsname\relax
  \def\url#1{\texttt{#1}}\fi
\expandafter\ifx\csname urlprefix\endcsname\relax\def\urlprefix{URL }\fi
\providecommand{\bibinfo}[2]{#2}
\providecommand{\eprint}[2][]{\url{#2}}

\bibitem[{\citenamefont{Yu et~al.}(1992)\citenamefont{Yu, McCaldin, and
  McGill}}]{Yu1992}
\bibinfo{author}{\bibfnamefont{E.}~\bibnamefont{Yu}},
  \bibinfo{author}{\bibfnamefont{J.}~\bibnamefont{McCaldin}}, \bibnamefont{and}
  \bibinfo{author}{\bibfnamefont{T.}~\bibnamefont{McGill}},
  \bibinfo{journal}{in \emph{Solid State Physics, Advances in Research and Applications}, vol.46, edited by {\bibfnamefont{H.}~\bibnamefont{Ehrenreich}} \bibnamefont{and} {\bibfnamefont{D.}~\bibnamefont{Turnbull}} (Academic Press, Inc., 1992)},
\bibinfo{pages}{pp. 1-146}.

\bibitem[{\citenamefont{Vurgaftman et~al.}(2001)\citenamefont{Vurgaftman,
  Meyer, and Ram-Mohan}}]{Vurgaftman2001}
\bibinfo{author}{\bibfnamefont{I.}~\bibnamefont{Vurgaftman}},
  \bibinfo{author}{\bibfnamefont{J.}~\bibnamefont{Meyer}}, \bibnamefont{and}
  \bibinfo{author}{\bibfnamefont{L.}~\bibnamefont{Ram-Mohan}},
  \bibinfo{journal}{J. Appl. Phys.} \textbf{\bibinfo{volume}{89}},
  \bibinfo{pages}{5815} (\bibinfo{year}{2001}).

\bibitem[{\citenamefont{Adachi et~al.}(2009)\citenamefont{Adachi, Capper,
  Kasap, and Willoughby}}]{Adachi2009}
\bibinfo{author}{\bibfnamefont{S.}~\bibnamefont{Adachi}},
  \bibinfo{journal}{in \emph{Properties of Semiconductor Alloys: Group-IV, III-V and
  II-VI Semiconductors}, edited by {\bibfnamefont{P.}~\bibnamefont{Capper}}, {\bibfnamefont{S.}~\bibnamefont{Kasap}}, \bibnamefont{and} {\bibfnamefont{A.}~\bibnamefont{Willoughby}} (Wiley Series in Materials for Electronic and
  Optoelectronic Applications, 2009)},
  \bibinfo{pages}{pp. 275-283}.

\bibitem[{\citenamefont{Perdew et~al.}(1996)\citenamefont{Perdew, Burke, and
  Ernzerhof}}]{Perdew1996}
\bibinfo{author}{\bibfnamefont{J.}~\bibnamefont{Perdew}},
  \bibinfo{author}{\bibfnamefont{K.}~\bibnamefont{Burke}}, \bibnamefont{and}
  \bibinfo{author}{\bibfnamefont{M.}~\bibnamefont{Ernzerhof}},
  \bibinfo{journal}{Phys. Rev. Lett.} \textbf{\bibinfo{volume}{77}},
  \bibinfo{pages}{3865} (\bibinfo{year}{1996}).

\bibitem[{\citenamefont{Heyd et~al.}(2003)\citenamefont{Heyd, Scuseria, and
  Ernzerhof}}]{Heyd2003}
\bibinfo{author}{\bibfnamefont{J.}~\bibnamefont{Heyd}},
  \bibinfo{author}{\bibfnamefont{G.~E.}~\bibnamefont{Scuseria}}, \bibnamefont{and}
  \bibinfo{author}{\bibfnamefont{M.}~\bibnamefont{Ernzerhof}},
  \bibinfo{journal}{J. Chem. Phys.} \textbf{\bibinfo{volume}{118}},
  \bibinfo{pages}{8207} (\bibinfo{year}{2003});
  \bibinfo{journal}{J. Chem. Phys.} \textbf{\bibinfo{volume}{124}},
  \bibinfo{pages}{219906} (\bibinfo{year}{2006}).

\bibitem[{\citenamefont{Moses and Van~de Walle}(2010)}]{Moses2010}
\bibinfo{author}{\bibfnamefont{P.~G.}~\bibnamefont{Moses}} \bibnamefont{and}
  \bibinfo{author}{\bibfnamefont{C.~G.}~\bibnamefont{Van~de Walle}},
  \bibinfo{journal}{Appl. Phys. Lett.} \textbf{\bibinfo{volume}{96}},
  \bibinfo{pages}{021908} (\bibinfo{year}{2010}).

\bibitem[{Wad()}]{Wadehra2010}
\bibinfo{journal}{See supplementary material at URL for bandgaps of
III-V binary and pseudobinary alloys, and details of the
special quasirandom structure.}

\bibitem[{\citenamefont{Nagao et~al.}(1997)\citenamefont{Nagao, Fujimoto,
  Gotoh, Fukushima, Takano, Ito, Koshihara, and Minami}}]{Nagao1997}
\bibinfo{author}{\bibfnamefont{S.}~\bibnamefont{Nagao}},
  \bibinfo{author}{\bibfnamefont{T.}~\bibnamefont{Fujimori}},
  \bibinfo{author}{\bibfnamefont{H.}~\bibnamefont{Gotoh}},
  \bibinfo{author}{\bibfnamefont{H.}~\bibnamefont{Fukushima}},
  \bibinfo{author}{\bibfnamefont{T.}~\bibnamefont{Takano}},
  \bibinfo{author}{\bibfnamefont{H.}~\bibnamefont{Ito}},
  \bibinfo{author}{\bibfnamefont{S.}~\bibnamefont{Koshihara}},
  \bibnamefont{and} \bibinfo{author}{\bibfnamefont{F.}~\bibnamefont{Minami}},
  \bibinfo{journal}{J. Appl. Phys.} \textbf{\bibinfo{volume}{81}},
  \bibinfo{pages}{1417} (\bibinfo{year}{1997}).

\bibitem[{\citenamefont{Waldrop et~al.}(1991)\citenamefont{Waldrop, Kraut,
  Farley, and Grant}}]{Waldrop1991}
\bibinfo{author}{\bibfnamefont{J.~R.}~\bibnamefont{Waldrop}},
  \bibinfo{author}{\bibfnamefont{E.~A.}~\bibnamefont{Kraut}},
  \bibinfo{author}{\bibfnamefont{C.~W.}~\bibnamefont{Farley}}, \bibnamefont{and}
  \bibinfo{author}{\bibfnamefont{R.~W.}~\bibnamefont{Grant}}, \bibinfo{journal}{J.
  Appl. Phys.} \textbf{\bibinfo{volume}{69}}, \bibinfo{pages}{372}
  (\bibinfo{year}{1991}).

\bibitem[{\citenamefont{Bohrer et~al.}(1993)\citenamefont{Bohrer, Krost, Wolf,
  and Bimberg}}]{Bohrer1993}
\bibinfo{author}{\bibfnamefont{J.}~\bibnamefont{Bohrer}},
  \bibinfo{author}{\bibfnamefont{A.}~\bibnamefont{Krost}},
  \bibinfo{author}{\bibfnamefont{T.}~\bibnamefont{Wolf}}, \bibnamefont{and}
  \bibinfo{author}{\bibfnamefont{D.}~\bibnamefont{Bimberg}},
  \bibinfo{journal}{Phys. Rev. B} \textbf{\bibinfo{volume}{47}},
  \bibinfo{pages}{6439} (\bibinfo{year}{1993}).

\bibitem[{\citenamefont{Patel et~al.}(1993)\citenamefont{Patel, Hafich,
  Robinson, and Menoni}}]{Patel1993}
\bibinfo{author}{\bibfnamefont{D.}~\bibnamefont{Patel}},
  \bibinfo{author}{\bibfnamefont{M.~J.}~\bibnamefont{Hafich}},
  \bibinfo{author}{\bibfnamefont{G.~Y.}~\bibnamefont{Robinson}}, \bibnamefont{and}
  \bibinfo{author}{\bibfnamefont{C.~S.}~\bibnamefont{Menoni}},
  \bibinfo{journal}{Phys. Rev. B} \textbf{\bibinfo{volume}{48}},
  \bibinfo{pages}{18031} (\bibinfo{year}{1993}).

\bibitem[{\citenamefont{Kim et~al.}(1995)\citenamefont{Kim, Cho, Choe, Jeong,
  and Lim}}]{Kim1995}
\bibinfo{author}{\bibfnamefont{K.~S}~\bibnamefont{Kim}},
  \bibinfo{author}{\bibfnamefont{Y.~H}~\bibnamefont{Cho}},
  \bibinfo{author}{\bibfnamefont{B.~D}~\bibnamefont{Choe}},
  \bibinfo{author}{\bibfnamefont{W.~G.}~\bibnamefont{Jeong}}, \bibnamefont{and}
  \bibinfo{author}{\bibfnamefont{H.}~\bibnamefont{Lim}},
  \bibinfo{journal}{Appl. Phys. Lett.} \textbf{\bibinfo{volume}{67}},
  \bibinfo{pages}{1718} (\bibinfo{year}{1995}).

\bibitem[{\citenamefont{Schneider et~al.}(1994)\citenamefont{Schneider, Jones,
  and Follstaedt}}]{Schneider1994}
\bibinfo{author}{\bibfnamefont{R.~P.}~\bibnamefont{Schneider}},
  \bibinfo{author}{\bibfnamefont{E.~D.}~\bibnamefont{Jones}}, \bibnamefont{and}
  \bibinfo{author}{\bibfnamefont{D.~M.}~\bibnamefont{Follstaedt}},
  \bibinfo{journal}{Appl. Phys. Lett.} \textbf{\bibinfo{volume}{65}},
  \bibinfo{pages}{587} (\bibinfo{year}{1994}).

\bibitem[{\citenamefont{Baldereschi et~al.}(1988)\citenamefont{Baldereschi,
  Baroni, and Resta}}]{Baldereschi1988}
\bibinfo{author}{\bibfnamefont{A.}~\bibnamefont{Baldereschi}},
  \bibinfo{author}{\bibfnamefont{S.}~\bibnamefont{Baroni}}, \bibnamefont{and}
  \bibinfo{author}{\bibfnamefont{R.}~\bibnamefont{Resta}},
  \bibinfo{journal}{Phys. Rev. Lett.} \textbf{\bibinfo{volume}{61}},
  \bibinfo{pages}{734} (\bibinfo{year}{1988}).

\bibitem[{\citenamefont{Nicklas and Wilkins}(2010)}]{Nicklas2010}
\bibinfo{author}{\bibfnamefont{J.~W.}~\bibnamefont{Nicklas}} \bibnamefont{and}
  \bibinfo{author}{\bibfnamefont{J.~W.}~\bibnamefont{Wilkins}},
  \bibinfo{journal}{Appl. Phys. Lett., under review}.

\bibitem[{\citenamefont{Blochl}(1994)}]{Blochl1994}
\bibinfo{author}{\bibfnamefont{P.~E.}~\bibnamefont{Blochl}},
  \bibinfo{journal}{Phys. Rev. B} \textbf{\bibinfo{volume}{50}},
  \bibinfo{pages}{17953} (\bibinfo{year}{1994}).

\bibitem[{\citenamefont{Kresse and Furthmuller}(1996)}]{Kresse1996}
\bibinfo{author}{\bibfnamefont{G.}~\bibnamefont{Kresse}} \bibnamefont{and}
  \bibinfo{author}{\bibfnamefont{J.}~\bibnamefont{Furthmuller}},
  \bibinfo{journal}{Phys. Rev. B} \textbf{\bibinfo{volume}{54}},
  \bibinfo{pages}{11169} (\bibinfo{year}{1996}).

\bibitem[{\citenamefont{Kresse and Joubert}(1999)}]{Kresse1999}
\bibinfo{author}{\bibfnamefont{G.}~\bibnamefont{Kresse}} \bibnamefont{and}
  \bibinfo{author}{\bibfnamefont{H.}~\bibnamefont{Joubert}},
  \bibinfo{journal}{Phys. Rev. B} \textbf{\bibinfo{volume}{59}},
  \bibinfo{pages}{1758} (\bibinfo{year}{1999}).

\bibitem[{\citenamefont{Paier et~al.}(2006)\citenamefont{Paier, Marsman,
  Hummer, Kresse, Gerber, and Angyan}}]{Paier2006}
\bibinfo{author}{\bibfnamefont{J.}~\bibnamefont{Paier}},
  \bibinfo{author}{\bibfnamefont{M.}~\bibnamefont{Marsman}},
  \bibinfo{author}{\bibfnamefont{K.}~\bibnamefont{Hummer}},
  \bibinfo{author}{\bibfnamefont{G.}~\bibnamefont{Kresse}},
  \bibinfo{author}{\bibfnamefont{I.~C.}~\bibnamefont{Gerber}}, \bibnamefont{and}
  \bibinfo{author}{\bibfnamefont{J.~G.}~\bibnamefont{Angyan}},
  \bibinfo{journal}{J. Chem. Phys.} \textbf{\bibinfo{volume}{124}},
  \bibinfo{pages}{154709} (\bibinfo{year}{2006}).

\bibitem[{\citenamefont{Zunger et~al.}(1990)\citenamefont{Zunger, Wei,
  Ferreira, and Bernard}}]{Zunger1990}
\bibinfo{author}{\bibfnamefont{A.}~\bibnamefont{Zunger}},
  \bibinfo{author}{\bibfnamefont{S.~-H.}~\bibnamefont{Wei}},
  \bibinfo{author}{\bibfnamefont{L.~G.}~\bibnamefont{Ferreira}}, \bibnamefont{and}
  \bibinfo{author}{\bibfnamefont{J.~E.}~\bibnamefont{Bernard}},
  \bibinfo{journal}{Phys. Rev. Lett.} \textbf{\bibinfo{volume}{65}},
  \bibinfo{pages}{353} (\bibinfo{year}{1990}).

\bibitem[{\citenamefont{van~de Walle et~al.}(2002)\citenamefont{van~de Walle,
  Asta, and Ceder}}]{Walle2002}
\bibinfo{author}{\bibfnamefont{A.}~\bibnamefont{van~de Walle}},
  \bibinfo{author}{\bibfnamefont{M.}~\bibnamefont{Asta}}, \bibnamefont{and}
  \bibinfo{author}{\bibfnamefont{G.}~\bibnamefont{Ceder}},
  \bibinfo{journal}{CALPHAD} \textbf{\bibinfo{volume}{26}},
  \bibinfo{pages}{539} (\bibinfo{year}{2002}).

\end{thebibliography}
\end{document}